\documentclass[10pt, twocolumn, superscriptaddress]{revtex4-1}

\usepackage{here}
\usepackage{amsmath}         
\usepackage{graphicx}
\usepackage{braket}         
 \usepackage{lettrine}
 \usepackage{color}
 \usepackage{times}

\begin{document}

\title{
Surface orbitronics: new twists from orbital Rashba physics
}    

\author{Dongwook Go}
\email{godw2718@postech.ac.kr}
\affiliation{Peter Gr\"unberg Institut and Institute of Advanced Simulation, Forschungszentrum J\"ulich and JARA, 52425 J\"ulich, Germany}
\affiliation{Department of Physics, Pohang University of Science and Technology, Pohang 37673, Korea}

\author{Jan-Philipp Hanke}
\author{Patrick M. Buhl}
\author{Frank Freimuth}
\author{Gustav Bihlmayer}
\affiliation{Peter Gr\"unberg Institut and Institute of Advanced Simulation, Forschungszentrum J\"ulich and JARA, 52425 J\"ulich, Germany}

\author{Hyun-Woo Lee}
\affiliation{Department of Physics, Pohang University of Science and Technology, Pohang 37673, Korea}

\author{Yuriy Mokrousov}
\author{Stefan Bl\"ugel}
\affiliation{Peter Gr\"unberg Institut and Institute of Advanced Simulation, Forschungszentrum J\"ulich and JARA, 52425 J\"ulich, Germany}

\date{\today}

\begin{abstract}
When the inversion symmetry is broken at a surface, spin-orbit interaction gives rise to spin-dependent  energy shifts $-$
a phenomenon which is known as the spin Rashba effect. Recently, it has been recognized that an orbital counterpart of the spin Rashba effect $-$ the orbital Rashba effect $-$ can be realized at surfaces even without spin-orbit coupling. Here, we propose a mechanism for the orbital Rashba effect based on $sp$ orbital hybridization, which ultimately leads to
the electric polarization of surface states. As a proof of principle, we show from first principles that this effect leads to chiral orbital textures in $\mathbf{k}$-space of the $\textup{Bi}\textup{Ag}_2$ monolayer.  In predicting the magnitude of the orbital moment arising from the orbital Rashba effect, we demonstrate the crucial role that the Berry phase theory plays for the magnitude and variation of the 
orbital textures. As a result, we predict a pronounced manifestation of various orbital effects at surfaces, and proclaim the orbital Rashba effect to be a key platform for surface orbitronics.

\end{abstract}

\maketitle
\date{\today}                 	      

\begin{figure*}
\includegraphics[angle=0, width=0.79\textwidth]{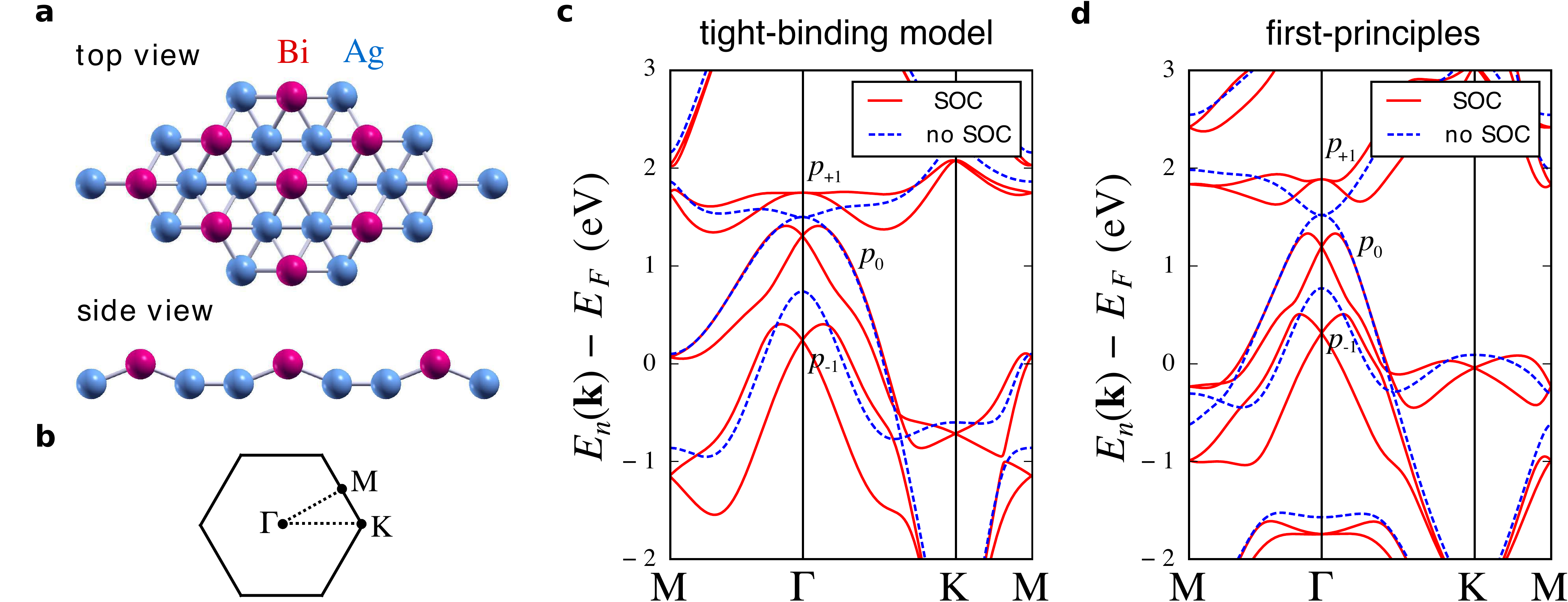}
\caption{\textbf{Crystal and electronic structure of $\textup{Bi}\textup{Ag}_2$.} \textbf{(a)} Top and side view of the crystal structure \cite{Xcrysden}. The combination of a strong surface potential gradient and a pronounced hybridization between Bi $6p$- and Ag $5s$-states gives rise to the orbital Rashba effect. \textbf{(b)} The hexagonal Brillouin zone and high-symmetry points. The tight-binding \textbf{(c)} and the first principles \textbf{(d)} band structures were calculated with and without taking into account spin-orbit coupling (SOC). States of distinct orbital symmetry are denoted as $p_{-1}$, $p_0$, and $p_{+1}$ following the convention of the main text. The tight-binding parameters were chosen so as to closely reproduce the first principles bandstructure.}
\end{figure*}

\noindent
\lettrine{T}he spin Rashba effect as a fundamental manifestation of the spin-orbit coupling (SOC) at surfaces has revolutionized the spintronics research, and served as
a foundation for a new field of spintronics rooted in relativistic effects $-$ {\it spin-orbitronics}~\cite{Rashba1960, Nitta1997, LaShell1996, Manchon2015}. As a consequence of the spin Rashba effect,
the surface states created as a result of SOC and surface potential gradient exhibit finite spin polarization, which forms chiral textures in reciprocal $\mathbf{k}$-space. This gives rise to a multitude of prominent phenomena such as Dzyaloshinskii-Moriya interaction~\cite{Dzyaloshinskii1957, Moriya1960, Lounis2012, Kim2013a, Kikuchi2016}, spin Hall effect~\cite{Sinova2004}, direct and inverse Edelstein effects~\cite{Edelstein1990, Freimuth2015}, quantum anomalous Hall effect~\cite{Chang2013}, and
current-induced spin-orbit torques~\cite{Miron2011, Garello2013, Kurebayashi2014}.

Only recently it has been realized that many of the spin-orbitronic effects can be rethought in terms of their ``parent" orbital analogs brought to the ``spin" level by SOC~\cite{Park2011,  Park2013, Hong2015, Kim2012, Park2012a, Bernevig2005, Kontani2009}. This leads to the notion of {\it orbitronics} as a promising branch of electronics which operates with the orbital degree of electrons  rather than their spin. The orbital moment (OM) of electrons in a solid, being an evasive quantity until very recently~\cite{Ceresoli2010, Lopez2012}, offers higher flexibility with respect to its magnitude and internal structure as compared to spin, and it is a key quantity for novel and promising effects in orbitronics such as gyrotropic magnetic effect~\cite{Zhong2016} and orbital Edelstein effect~\cite{Yoda2015}. The orbitronic analogue of the spin Rashba effect is the orbital Rashba effect (ORE),
which results in the emergence of chiral OM textures of the states in $\mathbf{k}$-space even without SOC, and manifests in the ``conventional" Rashba effect in the spin channel upon including SOC into the picture~\cite{Park2011, Park2013, Hong2015, Kim2012, Park2012a}. Indeed, circular dichroism measurements of $\textup{Au}(111)$ \cite{Kim2012} and $\textup{Bi}_2\textup{Se}_3$ \cite{Park2012a} surfaces are consistent with the prediction of the ORE forming non-trivial chiral OM textures in $\mathbf{k}$-space. On the other hand, quasi-particle interference experiments performed on another Rashba systems, the surface of $\textup{Pb}/\textup{Ag}(111)$ \cite{Kareh2014} and $\textup{Bi}/\textup{Ag}(111)$ \cite{Schirone2015}, demonstrate clear fingerprints of spin-orbital-flip scattering, which implies the vital role of the orbital degree of freedom of electrons for surface scattering.
As the spin Rashba effect played a key role in the development of spintronics, our understanding of the ORE and the ability to control its properties is pivotal for bringing orbitronics to the surface realm.

In this work, we introduce a mechanism for the ORE at surfaces, which originates in the orbital $sp$ hybridization and gives rise to pronounced ORE even without SOC in $sp$ surface alloys such as $\textup{Bi}\textup{Ag}_2$ monolayer. We elucidate that the $sp$ hybridization gives rise to the
orbital magnetoelectric coupling and intrinsically links the ORE  with electric polarization of states at the surface. Moreover, by referring to explicit first principles calculations, we demonstrate that including non-local effects \cite{Xiao2005, Thonhauser2005, Ceresoli2006, Shi2007, Xiao2010, Nikolaev2014, Hanke2016} can drastically enhance the magnitude of the ORE-driven orbital polarization of the surface states 
as compared to the $\mathbf{k}$-dependent OM obtained from simple atomic arguments.
We speculate that such an enhancement, which is particularly prominent in the vicinity of band crossings
in the surface electronic structure, can result in a giant magnitude of various orbitronics phenomena
such as orbital Hall effect \cite{Bernevig2005, Kontani2009}, gyrotropic magnetic effect \cite{Zhong2016}, and current-induced orbital magnetization \cite{Yoda2015}. This
marks the ORE as the most promising platform for realizing orbitronics at surfaces.

\

\noindent
\textbf{\large{Results}}

\noindent
\textbf{Tight-binding model.} 
We start the discussion of the ORE with tight-binding considerations. While previously 
only pure $p$- and $d$-orbital systems have been studied~\cite{Park2013, Petersen2000}, here we study monolayer of $sp$-derived surface alloys such as $\textup{Bi}\textup{Ag}_2$ \cite{Ast2007, Bian2013, Kareh2014}, which is different in that $sp$ hybridization is important. The tight-binding Hamiltonian can be generally written as:
\begin{eqnarray}
\label{eq:hopping_Hamiltonian}
H_\textup{hop}(\mathbf{k})
=
\begin{pmatrix}
H_p (\mathbf{k}) & h(\mathbf{k})
\\
h^\dagger (\mathbf{k}) & H_s (\mathbf{k})
\end{pmatrix},
\end{eqnarray}
where $H_{p(s)}(\mathbf{k})$ is a Hamiltonian in the subspace spanned by $p(s)$ orbitals, and $h(\mathbf{k})$ describes the effect of $sp$ hybridization. For a two-dimensional square 
lattice in the $xy$-plane with three $p$ orbitals and one $s$ orbital at each site, we choose the 
basis states as $\ket{\varphi_{n\mathbf{k}}}=({1}/{\sqrt{N}}) \sum_{\mathbf{R}} e^{i\mathbf{k}\cdot\mathbf{R}}\ket{\phi_{n\mathbf{R}}}$. Here, $\phi_{n\mathbf{R}}$ denotes the $n$-th ($n=p_x,p_y,p_z,s$) atomic orbital centered at the Bravais lattice $\mathbf{R}$, and $N$ is the number of lattice sites. In this basis, the hybridization assumes the
form:
\begin{eqnarray}
h (\mathbf{k}) = \left(  i\gamma_{sp}\sin (k_x a), i\gamma_{sp}\sin (k_y a),  V_z (\mathbf{k}) \right)^\mathrm{T} ,
\end{eqnarray}
where $\gamma_{sp}$ is the nearest-neighbor hopping amplitude between $p_{x(y)}$ and $s$ orbitals, $a$ is the lattice constant, $V_z (\mathbf{k}) = e\mathcal{E}_z \bra{\varphi_{p_z\mathbf{k}}} z \ket{\varphi_{s\mathbf{k}}}$ describes the effect of $sp_z$ hybridization by a surface potential gradient $\mathcal{E}_z$, and $e>0$ is the elementary positive charge. The effect of SOC is included into the Hamiltonian as 
$H_\textup{soc} = \lambda_{\textup{soc}} \hat{\mathbf{L}} \cdot \hat{\mathbf{S}}$,
where $\lambda_{\textup{soc}}$ is the spin-orbit strength, $\hat{\mathbf{S}}$ is the operator of spin,
and 
\begin{eqnarray}
&\hat{\mathbf{L}} =
\sum_\mathbf{R}
\hat{\mathbf{L}}_\mathbf{R},
\nonumber
\\
&\hat{\mathbf{L}}_\mathbf{R}
=
\sum_{nm}
\ket{\phi_{n\mathbf{R}}} \bra {\phi_{n\mathbf{R}}} ( \mathbf{r-R}) \times \mathbf{p} \ket{\phi_{m\mathbf{R}}}\bra{\phi_{m\mathbf{R}}}
\label{eq:atom-centerd approximation}
\end{eqnarray}
is the representation of the atomic contribution of the orbital angular momentum operator \cite{Bihlmayer2006}. Here, $\mathbf r$ and $\mathbf p$ denote the canonical position and momentum operators, respectively. Although the general conclusions that we draw from the tight-binding model do not depend on the exact choice of the hopping parameters, they can be easily tuned such that the tight-binding band structure closely resembles the first principles one of the $\textup{Bi}\textup{Ag}_2$ monolayer (Fig.~1).

\

\noindent
\textbf{Orbital Rashba effect.} 
Neglecting for the moment the effect of SOC and assuming $|H_p(\mathbf{k})- H_s(\mathbf{k})|\gg |h(\mathbf{k})|$ as is the case for $\textup{Bi}\textup{Ag}_2$ in the long-wavelength limit of $\mathbf{k}\rightarrow 0$, we can perturbatively downfold the $h(\mathbf{k})$ term to arrive at an effective Hamiltonian (see Supplementary Information):
\begin{eqnarray}
H_\textup{eff} (\mathbf{k}) 
=
\begin{pmatrix}
H_{p,\textup{eff}} (\mathbf{k}) & 0
\\
0 & H_{s,\textup{eff}} (\mathbf{k})
\end{pmatrix},
\end{eqnarray}
where $H_{p,\textup{eff}}(\mathbf{k}) = H_p (\mathbf{k}) + H_\textup{OR}(\mathbf{k})$. The expression
\begin{equation}
\label{eq:orbital Rashba Hamiltonian}
H_\textup{OR} (\mathbf{k}) = \frac{\alpha_\textup{OR}}{\hbar}\, \hat{\mathbf{L}} \cdot (\hat{\mathbf{z}}\times\mathbf{k})
\end{equation}
is known as the {\it orbital Rashba Hamiltonian} since it resembles the conventional spin Rashba Hamiltonian with the orbital angular momentum operator replacing that of spin. Remarkably, the combined effect of $sp$ hybridization and surface potential gradient is concisely described by the orbital Rashba Hamiltonian within the subspace spanned by $p$ orbitals. Thus, the orbital Rashba physics arises not from the electron's spin but from the orbital degrees of freedom even without SOC. In analogy to the Rashba constant of the conventional spin Rashba model, the parameter 
\begin{eqnarray}
\label{eq:orbital Rashba constant}
\alpha_\textup{OR}
=
\left.
\frac{\eta\, a\,\gamma_{sp}\,  V_z (\mathbf{k}) }{\Delta E_{sp}(\mathbf{k})}
\right|_{\mathbf{k}=0}
\end{eqnarray}
is called the {\it orbital Rashba constant}, with $\Delta E_{sp}(\mathbf{k})$ denoting the energy gap between $s$- and $p$-derived bands, and $\eta\sim 1$ being a parameter dependent on the lattice structure. Using the specific tight-binding model parameters for $\textup{Bi}\textup{Ag}_2$ (see Supplementary Information), we estimate the orbital Rashba constant to be about $ 1\ \textup{eV}\cdot\mathrm{\AA}$. From Eq.~(\ref{eq:orbital Rashba constant}) it is clear that the ORE roots in the $sp$ orbital hybridization, and that the strength of the orbital Rashba effect is directly proportional to the value of the surface potential
gradient associated with the buckling of $\textup{Bi}$ atoms in $\textup{Bi}\textup{Ag}_2$. This implies that the desired properties of the ORE can be designed by controlling the band hybridization via chemical and structural engineering.

\

\noindent
\textbf{Crystal field splitting.} In contrast to the conventional Rashba effect, the ORE is very sensitive to the crystal field splitting (CFS), which quenches the OM. In the absence of the CFS,~that is,~when $\Delta_\textup{CFS} = E_{p_{x(y)}} - E_{p_z} = 0$ with $E_{p_{x(y)}}, E_{p_z}$ as corresponding energy eigenvalues of $p_{x(y)}$- and $p_z$-derived states at $\mathbf{k}=0$, the expectation value of the OM is given by 
\begin{eqnarray}
\mathbf{m}^\textup{ACA}_{l} (\mathbf{k})
&=&
-\frac{\mu_B}{\hbar} \sum_\mathbf{R}
\bra{\psi_{p_l\mathbf{k}}} \hat{\mathbf{L}}_\mathbf{R} \ket{\psi_{p_l\mathbf{k}}} 
=
 \mu_B l \ \hat{\mathbf{k}} \times \hat{\mathbf{z}},
 \nonumber
 \\
\label{eq: orbital chirality}
\end{eqnarray} 
within the so-called \emph{atom-centered approximation} for the OM, which takes into account only intra-atomic contributions. Here, $\hat{\mathbf{z}}$ and $\hat{\mathbf{k}}$ are unit vectors along the $z$-axis and and vector $\mathbf{k}$, respectively, $\psi_{p_l\mathbf{k}}$ is the $p_l$-derived eigenstate of $H_{p,\textup{eff}}(\mathbf{k})$, and $l=\{-1, 0, +1 \}$ is the angular momentum quantum number with respect to the quantization axis $\hat{\mathbf{z}} \times \hat{\mathbf{k}}$. In the vicinity of the Fermi energy, the $p$-derived bands can thus be denoted in terms of their dominant orbital character as $p_{-1}$, $p_0$, and $p_{+1}$ in the order of increasing energy (Fig.~1\textbf{c}-\textbf{d}). The corresponding $\mathbf{k}$-dependent eigenenergies are $E_{-1} (\mathbf{k})$, $E_{0}(\mathbf{k})$, and  $E_{+1}(\mathbf{k})$. The associated $\mathbf{k}$-linear orbital-dependent energy splitting arising due to the orbital Rashba term in $H_{p,\textup{eff}}(\mathbf{k})$ amounts to $\Delta E_\textup{OR} = l \alpha_\textup{OR}|\mathbf{k}|$ in the long-wavelength limit. However, in the presence of the CFS, although its direction remains intact, the expectation value of the OM is reduced by a factor of 
${2 \alpha_\textup{OR}}|\mathbf{k}|/{\Delta_\textup{CFS}}$, when assuming 
$|\Delta_\textup{CFS}| \gg |\alpha_\textup{OR}\mathbf{k}|$.
For this reason, orbital-dependent energy splitting appears from the second order in $\mathbf{k}$ if $\Delta_{\textup{CFS}} \neq 0$, which can be seen from Eq. (\ref{eq:orbital Rashba Hamiltonian}).

\

\noindent
\textbf{Relation to electric polarization.} Consider an orbital-coherent state ${\psi_{p_l\mathbf{k}}}$, that is, an eigenstate
 of $H_\textup{hop}(\mathbf{k})$ with $\Delta_\textup{CFS}=0$ which exhibits a quantized value of OM (e.g., due to an applied
 in-plane magnetic field or the ORE):
\begin{eqnarray}
\mathbf{m}^\textup{ACA}_{l} (\mathbf{k})
= \hbar l\, \hat{\mathbf{n}}_\mathbf{k},
\label{eq:temp1}
\end{eqnarray}
where $l=\{-1,0,+1\}$ is the orbital angular momentum quantum number with respect to the direction of some in-plane unit vector $\hat{\mathbf{n}}_\mathbf{k}$. While in previous studies the electric polarization within the ORE was introduced phenomenologically~\cite{Park2011, Hong2015}, one can show from perturbation theory arguments that such an orbital-coherent state naturally exhibits electric polarization perpendicular to the surface plane:
\begin{eqnarray}
P^z_l (\mathbf{k}) 
&=& -e \bra{\psi_{p_m\mathbf{k}}} z \ket{\psi_{p_m\mathbf{k}}} 
=
\frac{ \chi_\textup{OR}}{\mu_B} 
\mathbf{m}_l^\textup{ACA} (\mathbf{k})
\cdot
(\mathbf{k}\times \hat{\mathbf{z}}), 
\nonumber
\\
\label{eq:polarization}
\end{eqnarray}
where the parameter
\begin{equation}
\label{eq:orbital_Rashba_susceptibility}
\chi_\textup{OR} 
=
\left.
\frac{\eta\, a\,e\,\gamma_{sp}\bra{\varphi_{p_z\mathbf{k}}} z \ket{\varphi_{s\mathbf{k}}}}{\Delta E_{sp}(\mathbf{k})}
\right|_{\mathbf{k}=0}
\end{equation}
is what we call the {\it orbital Rashba susceptibility}. The structure of $\chi_\textup{OR}$ reflects that the electric polarization arises from the $sp$ hybridization (Fig. 2\textbf{a}). From Eq.~(\ref{eq:orbital Rashba constant}) it is clear that $\alpha_\textup{OR} = \mathcal{E}_z \chi_\textup{OR}$, and that the orbital Rashba Hamiltonian as given by Eq. (\ref{eq:orbital Rashba Hamiltonian}) can be understood as a dipole coupling between an electric field at the surface with the operator of electric polarization, $H_\textup{OR} = -\mathcal{E}_z \hat{P}_z (\mathbf{k})$, where $\hat{P}_z(\mathbf{k}) = -(\chi_\textup{OR}/\hbar) \,\mathbf{k}\times\hat{\mathbf{L}}$ (Fig. 2\textbf{b}). Physically, it means that within the ORE the non-trivial orbital texture of states in $\mathbf{k}$-space arises 
so as to gain maximal energy by the interaction of the states' polarization with the surface electric field. Thus,
the ORE can be seen as a consequence of {\it orbital magnetoelectric coupling} at surfaces.

\begin{figure}[t!]
\includegraphics[angle=0, width=0.45\textwidth]{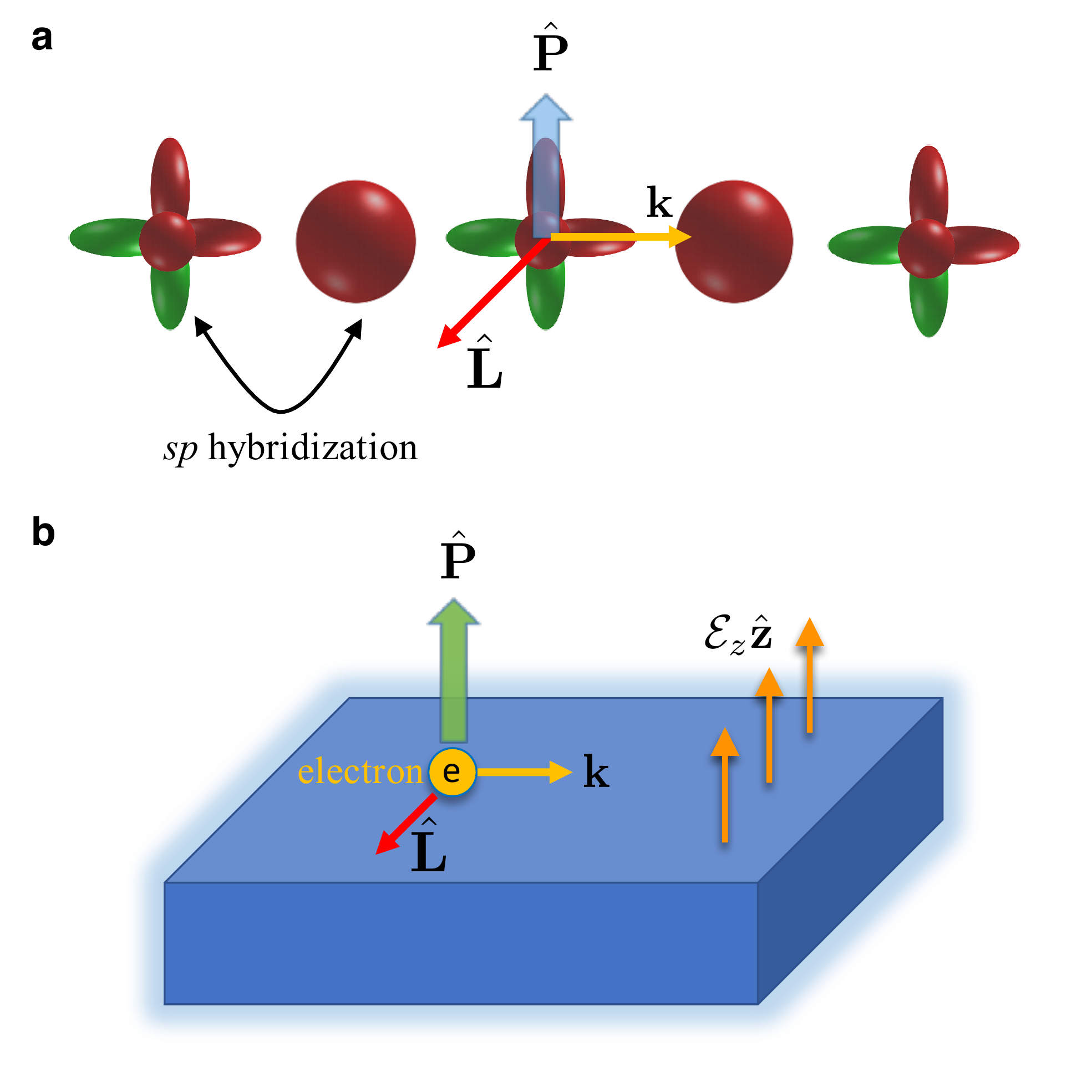}
\caption{
\textbf{Orbital Rashba effect in $sp$-alloys.} \textbf{(a)} After preparing the system in a state with non-vanishing orbital angular momentum $\hat{\mathbf L}$, for example, by applying an in-plane magnetic field, the hybridization between $s$ and $p$ orbitals generates an electric polarization $\hat{\mathbf P}$ according to Eq.~\eqref{eq:polarization}. \textbf{(b)} The electric dipole coupling of this electric polarization to the surface potential gradient $\mathcal E_z \hat{\mathbf z}$ [Eq.~\eqref{eq:orbital Rashba Hamiltonian}] leads to the formation of an orbital texture in reciprocal space, which is known as orbital Rashba effect.
}
\end{figure}

\begin{figure}[ht]
\includegraphics[angle=0, width=0.46\textwidth]{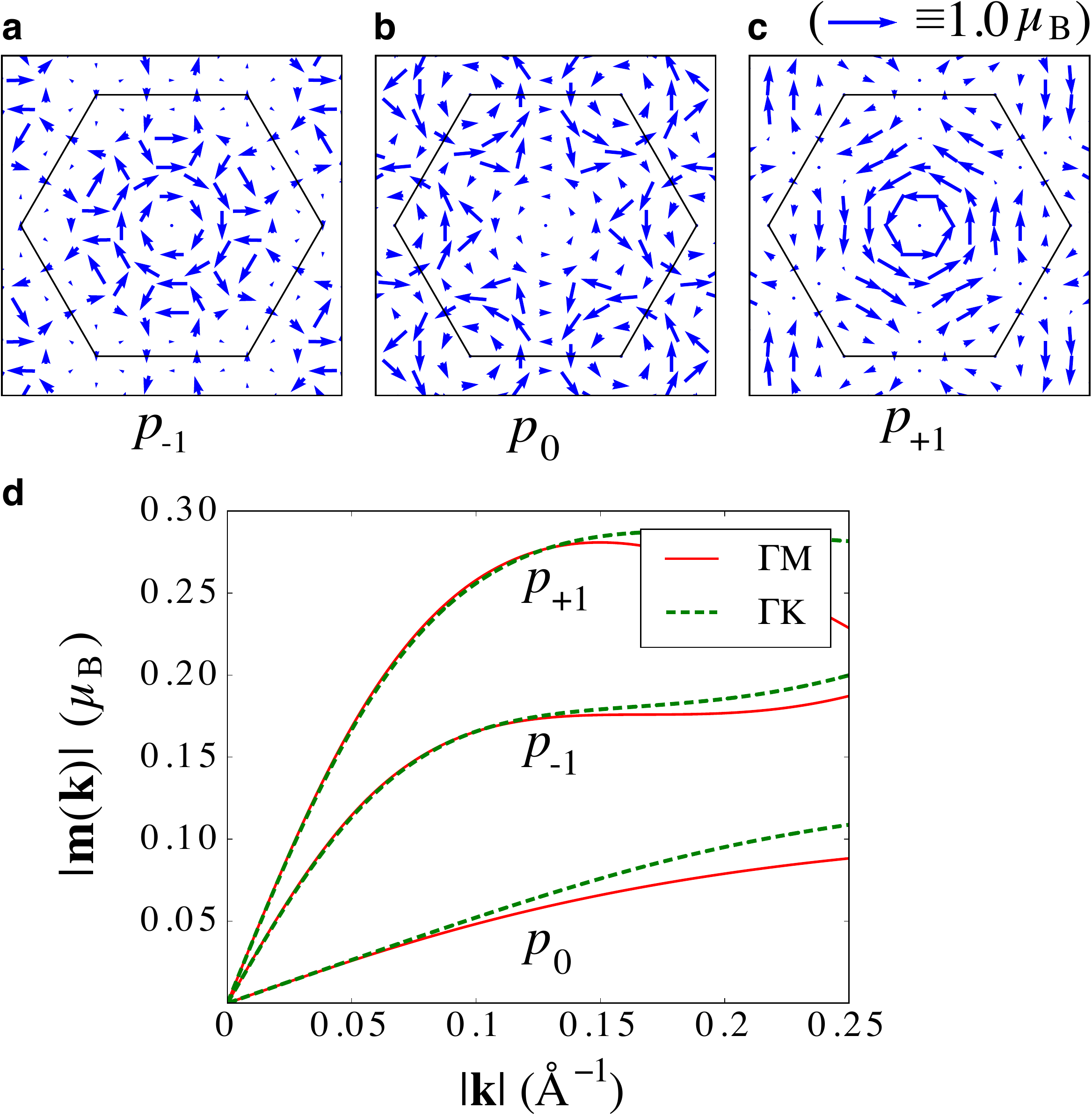}
\caption{\textbf{Orbital texture of $p$-derived bands in $\textup{Bi}\textup{Ag}_2$ without spin-orbit coupling.}
\textbf{(a)}-\textbf{(c)} Arrows indicate the first principles values for the $\mathbf k$-resolved in-plane orbital moment (OM) of the $p$-derived bands within the atom-centered approximation. The hexagonal Brillouin zone of $\textup{Bi}\textup{Ag}_2$ is indicated by thin black lines. In the vicinity of the $\Gamma$-point, the chirality of the OM texture is $+1$, $0$, and $-1$ for $p_{-1}$, $p_{0}$, and $p_{+1}$, respectively. \textbf{(d)} The magnitude of the in-plane OM $|\mathbf{m}_n (\mathbf{k})|$ of the $p$-derived bands along the high-symmetry lines $\Gamma \mathrm{M}$ and $\Gamma \mathrm{K}$.
}
\end{figure}

\begin{figure*}[ht!]
\includegraphics[angle=0, width=1.0\textwidth]{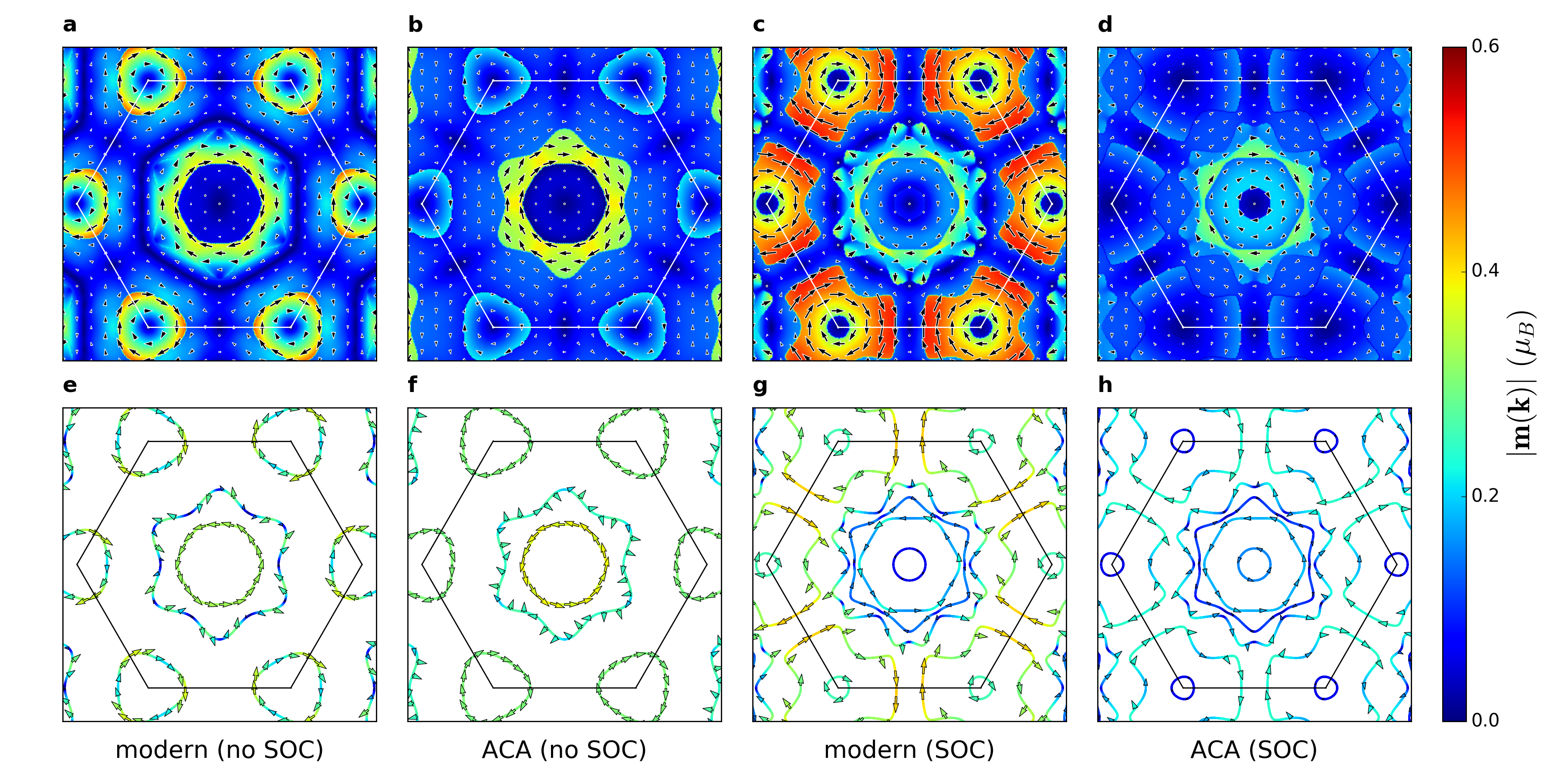}
\caption{\textbf{In-plane orbital moment (OM) in $\textup{Bi}\textup{Ag}_2$ from first principles.}
\textbf{(a)}-\textbf{(d)} Summing up the individual contributions of all occupied bands below the Fermi energy, we obtain the distribution of the in-plane OM $\mathbf{m} (\mathbf{k})$ in $\mathbf{k}$-space. \textbf{(e)}-\textbf{(h)} The in-plane OM directly at the Fermi surface, which is important to orbital magnetoelectric response. In all cases, colors represent the magnitude $|\mathbf{m} (\mathbf{k}) |$, arrows indicate the in-plane direction of the OM with the arrow size proportional to $|\mathbf{m} (\mathbf{k}) |$, and thin lines mark the Brillouin zone. The results of both Berry phase theory (modern) and atom-centered approximation (ACA) are shown with and without taking into account spin-orbit coupling (SOC). The in-plane OM within the two approaches deviates drastically around the $\mathrm{K}$-point, and the modern theory predicts overall a richer structure.}
\end{figure*}

\

\noindent
\textbf{Atom-centered approximation from first principles.} In order to verify our predictions in a realistic system, we evaluate the $\mathbf{k}$- and band-resolved value of the OM in $\textup{Bi}\textup{Ag}_2$ from first principles, which is given by $\mathbf{m}^\textup{ACA}_n (\mathbf{k})=-(\mu_B/\hbar)\sum_{\mu}\bra{\psi_{n\mathbf{k}}} \mathbf r_\mu\times \mathbf{p} \ket{\psi_{n\mathbf{k}}}_\mu$ in ACA. Here, $\psi_{n \mathbf k}$ is an eigenstate, $\mathbf r_\mu$ is the position operator relative to atom $\mu$, the summation is performed over all atoms in the lattice, and the real-space integration is restricted locally to \emph{atom-centered muffin-tin spheres}. Computed in such a way OM is a direct generalization
to the first principles framework of the OM computed in ACA from tight-binding.
In Fig. 3$\mathbf{a}$-$\mathbf{c}$, we show the distribution of the OM in $\mathbf{k}$-space for the $p$-derived bands $p_{-1}$, $p_0$, and $p_{+1}$ in absence of SOC, and observe clockwise ($-1$), zero, and counter-clockwise ($+1$) type of chiral behavior of this distribution around the $\Gamma$-point,  respectively. The magnitude of the OM without SOC along different high-symmetry lines is shown in Fig. 3$\textbf{d}$. While the OM reaches as much as $0.27 \mu_B$ for the $p_{+1}$ band in the middle of the Brillouin zone, we can use its behaviour with $\mathbf k$ in the vicinity of the $\Gamma$-point to estimate the magnitude of the orbital Rashba constant $\alpha_\textup{OR}$. First, we note that a small but finite value of the slope in the OM of $p_0$ as well as visible differences in the behavior of the OM of $p_{-1}$ and $p_{+1}$ bands can be observed, in contrast to our prediction Eq.~\eqref{eq:temp1}. The reason for the discrepancy lies in a non-zero crystal field splitting, which we can estimate from Fig. 1$\textbf{d}$ to be $\Delta_\textup{CFS} \approx 0.76\ \textup{eV}$. Taking this into account, the orbital Rashba constant of $p_{-1}$ and $p_{+1}$ bands amounts to $ 0.96\ \textup{eV}\cdot\textup{\AA}$ and $1.38\ \textup{eV}\cdot\textup{\AA}$, respectively. Similarly, we also find orbital chiralities near $\mathrm{K}$-point. Investigating further the influence of SOC, we found no qualitative changes in the distribution of the OM in the Brillouin zone. However,
we clearly see that the resulting ``Rashba" 
{\it spin} texture in reciprocal space, emerging upon including SOC, is strictly 
bound to the local direction of the OM. Details on these results are provided in the Supplementary Information. Finally, we display in Figs.~4\textbf{f} and ~4\textbf{g} the chirality of the OM at the Fermi surface both with and without SOC taken into consideration, respectively.

\

\noindent
\textbf{Berry phase theory in films.} The primary manifestation of the ORE in solids is the $\mathbf{k}$- and band-dependent generation of finite OM. For understanding this fundamental effect and predicting its magnitude, it is crucial to evaluate the magnitude of the OM properly without assuming any approximations such as the ACA. Recently, it was shown that a rigorous treatment of OM in solids within the complete Berry phase description \cite{Xiao2005, Thonhauser2005, Ceresoli2006, Shi2007} naturally accounts for non-local effects \cite{Nikolaev2014}. Thereby, theoretical estimations of OM in magnetic materials of various nature have significantly improved~\cite{Ceresoli2010, Lopez2012, Hanke2016}.

Since the ORE manifests in the generation of in-plane OM, following the procedure of Ref. \cite{Shi2007}, we extended the previous formulation to the case of in-plane components of OM of a film finite along the $z$-axis. The expression for the in-plane components of the OM  is given by 
\begin{eqnarray}
& &
\mathbf{m}_n^\textup{mod} (\mathbf{k})
=
\nonumber
\\
\nonumber
& & 
\frac{e}{\hbar}
\nonumber
\textup{Re}
\left[
\bra{ u_{n\mathbf{k}}}
( z \hat{\mathbf{z}} )
\times
\left\{
H(\mathbf{k}) + E_n (\mathbf{k}) -  2 E_F
\right\}
\ket{\partial_{\mathbf{k}} u_{n\mathbf{k}}}
\right],
\nonumber
\\
\label{eq:modern}
\end{eqnarray}
where $u_{n\mathbf{k}}$ is an eigenstate of the lattice-periodic Hamiltonian $H(\mathbf{k})$ with the eigenvalue $E_n(\mathbf{k})$, and $E_F$ is the Fermi energy. Equation~\eqref{eq:modern} can be further decomposed into the so-called local circulation term $\mathbf{m}_n^\textup{LC}(\mathbf{k}) 
=
(e/\hbar )\textup{Re} \left[ \bra{ u_{n\mathbf{k}}} (z \hat{\mathbf{z}}) \times \left\{ H(\mathbf{k}) - E_F \right\}  \ket{\partial_{\mathbf{k}} u_{n\mathbf{k}}} \right]
$, and the itinerant circulation term $\mathbf{m}_n^\textup{IC}(\mathbf{k}) 
=
(e/\hbar )  \left\{ H(\mathbf{k}) - E_F \right\} \textup{Re} \left[ \bra{ u_{n\mathbf{k}}} (z \hat{\mathbf{z}})\times   \ket{\partial_{\mathbf{k}} u_{n\mathbf{k}}} \right]$. The latter expression is connected to the {\it projected} Berry curvature
\begin{eqnarray}
\label{eq:Berry curvature}
\Omega_{x(y)}^n (\mathbf{k}) 
=
\textup{Re}
\left[
\bra{\partial_{k_{x(y)}} u_{n\mathbf{k}}} z \ket{u_{n\mathbf{k}}}
\right],
\end{eqnarray}
which is closely related to the well-known Berry curvature in bulk systems by formally replacing $z$ with $i\partial_{k_z}$.

Inserting the model orbital Rashba Hamiltonian, Eq.~\eqref{eq:orbital Rashba Hamiltonian}, directly into Eq. (\ref{eq:modern}), we find that
\begin{equation}
\label{eq: mod_vs_aca} 
\mathbf{m}_l^\textup{mod} (\mathbf{k})
= -\chi_\textup{OR}\,\frac{(E_{p_{x(y)}} + E_{p_z} - 2E_F)}{2\mu_B\hbar}
 \cdot 
\mathbf{m}_l^\textup{ACA} (\mathbf{k})
\end{equation}
in the long-wavelength limit, where $l$ is the angular-momentum quantum number defined in Eq. (\ref{eq: orbital chirality}). Assuming the typical values $\mathcal{E}_z \sim 0.1\ \mathrm{eV}/\mathrm{\AA}$ and $(E_{p_{x(y)}} + E_{p_z} - 2E_F) \sim 1\ \mathrm{eV}$, we find not only that ACA and modern OM exhibit the same chirality of the distribution, but also that $\mathbf{m}_l^\textup{mod} (\mathbf{k}) \sim \mathbf{m}_l^\textup{ACA} (\mathbf{k})$ in the vicinity of the $\Gamma$-point within the model analysis.

\

\noindent
\textbf{Berry phase theory from first principles.} To investigate whether significant differences between the ACA and modern treatment of OM arise in a realistic situation, we evaluate from first principles $\mathbf{m}_n^\textup{ACA} (\mathbf{k})$ and $\mathbf{m}_n^\textup{mod} (\mathbf{k})$. In Fig. 4, we show the OM distribution evaluated from the modern theory (Figs. 4\textbf{a}, 4\textbf{c}) and ACA (Figs. 4\textbf{b}, 4\textbf{d}), where the individual contributions of all occupied bands were summed up for each $\mathbf{k}$ point. In this figure, the distribution of the OM within the modern approach and ACA is similar around the $\Gamma$-point both in magnitude and overall distribution,
in accordance with our model considerations. Near the $\textup{K}$-point, however, the distribution of the modern OM without SOC deviates significantly from the ACA result. If SOC is taken into account, the difference between the two approaches is even more drastic as it amounts to one order of magnitude. In Fig. 4, the visible discontinuity of the OM distribution occurs along the Fermi surface lines (Figs. 4\textbf{e}-4\textbf{h}). Directly at the Fermi surface, the itinerant contribution to the OM in the modern theory vanishes and we may restrict ourselves to visualizing in Figs.~4\textbf{e} and 4\textbf{g} the local circulation from the modern theory, without and with SOC, respectively. In contrast, within ACA there is no such decomposition of the Fermi-surface contribution (Figs.~4\textbf{f}, 4\textbf{h}). The OM of Fermi-surface states plays a crucial role in various orbital magnetoelectric phenomena~\cite{Zhong2016, Yoda2015}.

Within the Berry phase theory, one of the most remarkable features of the OM in bulk is its correlation with the Berry curvature in $\mathbf{k}$-space, which often exhibits a spiky behavior in the vicinity of band crossings~\cite{Xiao2010}. In our formalism for the in-plane OM in thin films as expressed by Eq.~(\ref{eq:modern}), the projected Berry curvature, Eq.~\eqref{eq:Berry curvature}, is a key ingredient. It behaves similarly to the conventional Berry curvature in that it can exhibit singular behavior at band crossings as a consequence of the rapid variation of the wave functions with $\mathbf k$. Following this spirit, we also seek for such a spiky behavior in the local circulation $\mathbf{m}^\textup{LC}(\mathbf{k})$ by studying the OM in the vicinity of the band crossing in the electronic structure of $\textup{Bi}\textup{Ag}_2$ which is close to the $\mathrm{M}$-point and about 0.2\,eV below the Fermi level. Setting $E_F$ in Eq.~\eqref{eq:modern} to the energy of the crossing and treating all bands below this energy as occupied, we observe a singular behavior of the OM magnitude within the modern theory in the vicinity of the crossing point (Fig. 5\textbf{a}). This behavior can be directly correlated with sizable OM contributions of those states which constitute the band crossing. At the point of singularity, the modern-theory OM is purely due to the local circulation and it reaches as much as $1.01\,\mu_B$ in magnitude, while the ACA predicts a tiny value of 
$0.03\,\mu_B$. Remarkably, both ACA and modern theory agree qualitatively in their prediction of the behavior of the OM away from the band crossing (see Supplementary Information). 

\begin{figure*}[ht!]
\includegraphics[angle=0, width=0.8\textwidth]{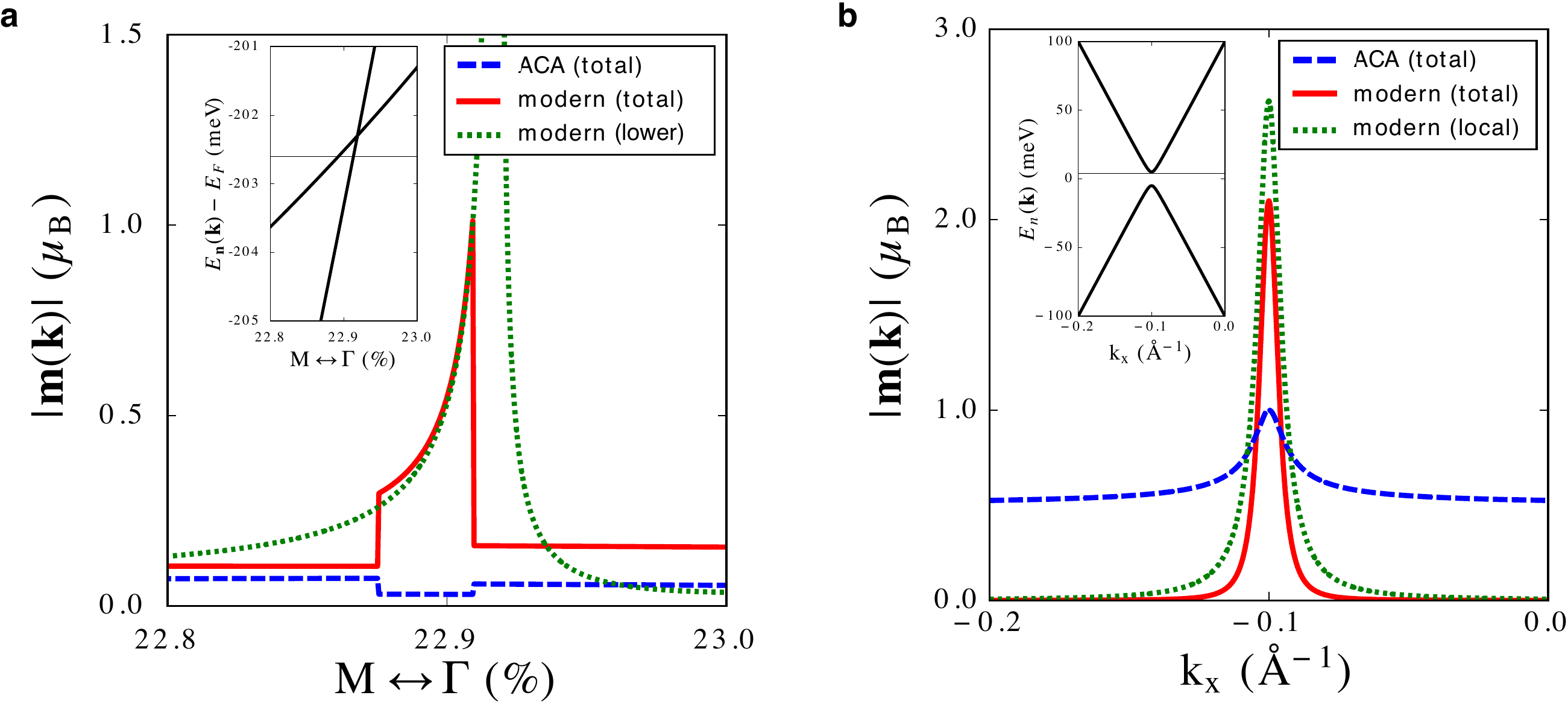}
\caption{\textbf{In-plane orbital moment (OM) near band crossings in $\textup{Bi}\textup{Ag}_2$.} \textbf{(a)} First principles results for the total in-plane OM near the band crossing, which is shown as an inset. All states below the energy indicated by the black horizontal line were considered as occupied in the band summation of the OM. While the ACA value is insensitive to the presence of band crossings, the Berry phase theory predicts a gigantic enhancement of the local circulation of the OM in the vicinity of such points. In particular, the band-resolved contributions (shown here for the lower band) display singular behavior due to the rapid variation of the wave function near the crossing. \textbf{(b)} Similar conclusions can be drawn from tight-binding model calculations assuming a Dirac-like dispersion (inset). Whereas the ACA prediction is strictly bound by $\mu_B$ within the band gap, pronounced values of the in-plane OM $|\mathbf{m} (\mathbf{k})|$ can be reached in the modern theory. Likewise, the corresponding Fermi-surface contribution is strongly enhanced in the gap region.}
\end{figure*}


\

\noindent
\textbf{Two-band model near the band crossing.} To study the aforementioned behavior of the OM near the band crossing from the model point-of-view, we consider the following two-band Hamiltonian:
\begin{eqnarray}
H (\mathbf{k}) = \mathbf{d} (\mathbf{k}) \cdot \boldsymbol{\sigma},
\label{h}
\end{eqnarray}
where $\boldsymbol{\sigma}$ is the vector of Pauli matrices reflecting the orbital degree of freedom of two chosen basis  states: $\ket{\varphi_{s,\mathbf{k}}}$, and 
$\ket{\varphi_{p,\mathbf{k}}}
=
(1/\sqrt{2}) \ket{\varphi_{p_x,\mathbf{k}}}
-
(i/\sqrt{2}) \ket{\varphi_{p_z,\mathbf{k}}}
$. 
We assume that $d_x(\mathbf{k}) = 0$, $d_y (\mathbf{k}) = \gamma_{sp}k_x a + V_z$, and $d_z (\mathbf{k}) = \delta$, where $a$ is the lattice constant, $\delta$ is the energy gap of the massive Dirac-like dispersion along $k_x$, $\gamma_{sp}$ is the nearest-neighbor hopping amplitude between $s$ and $p_x$ orbitals, and $V_z = -(e \mathcal{E}_z/\sqrt{2}) \bra{\varphi_{s,\mathbf{k}}}  z \ket{\varphi_{p_z,\mathbf{k}}}$ expresses the breaking of inversion symmetry. The band structure of this Hamiltonian is shown in Fig. 5$\textbf{a}$, where we assumed $a=5\ \textup{\AA}$, $V_z = 0.1\ \textup{eV}$, $\gamma_{sp} = 1 \ \textup{eV}$, and $\delta = 5\ \textup{meV}$. We represent the position operator $z$ in the specified basis as 
\begin{eqnarray}
z &=&
\begin{pmatrix}
\bra{\varphi_{s,\mathbf{k}}} z \ket{\varphi_{s,\mathbf k}} & \bra{\varphi_{s,\mathbf{k}}} z \ket{\varphi_{p,\mathbf k}} \\
\bra{\varphi_{p,\mathbf{k}}} z \ket{\varphi_{s,\mathbf k}} & \bra{\varphi_{p,\mathbf{k}}} z \ket{\varphi_{p,\mathbf k}}
\end{pmatrix}
\nonumber
\\
&=&
\mathbf{p}(\mathbf{k}) \cdot \boldsymbol{\sigma},
\label{p}
\end{eqnarray}
where $p_x = p_z = 0$ and $p_y = -(1/\sqrt{2}) \bra{\varphi_{s,\mathbf{k}}} z \ket{\varphi_{p_z,\mathbf{k}}}$.
Taking into account the representation (\ref{p}) and applying Eq.~(\ref{eq:modern}), we obtain an expression for the in-plane OM from the modern theory for the general two-band Hamiltonian (\ref{h}):
\begin{eqnarray}
\mathbf{m}_\pm^\textup{mod} (\mathbf{k})
&=&
\pm \frac{e}{\hbar} E_F 
\left[
\left(\hat{\mathbf{z}}\times \partial_{\mathbf{k}}\right) \hat{\mathbf{d}} (\mathbf{k})
\right] 
\cdot
\mathbf{p} (\mathbf{k}), 
\label{eq: two band 1}
\\
&\approx &
 - \frac{E_F}{\hbar}
\left(\hat{\mathbf{z}}\times \partial_{\mathbf{k}}\right) {P}_z^\pm (\mathbf{k})
\label{eq:two band 2}
\end{eqnarray}
where $\hat{\mathbf{d}} (\mathbf{k})$ is the direction of $\mathbf{d}(\mathbf{k})$ and the ``$+$" (``$-$") sign stands for upper (lower) band. In the second line, owing to the fact that  in our model $(\hat{\mathbf{z}}\times\partial_\mathbf{k})\mathbf{p}(\mathbf{k}) \approx 0$, we related the modern OM
to the derivative of ${P}_z^\pm (\mathbf{k}) = \pm (-e) [\hat{\mathbf{d}}(\mathbf{k})\cdot\mathbf{p}(\mathbf{k})]$, that is, the $z$-component of the electric polarization of the upper (lower) band as defined in Eq.~(\ref{eq:polarization}). 

From this generic expression we clearly observe  that the origin of the non-vanishing OM within the modern
theory lies in the gradient of the electric polarization in reciprocal space. Using the parameters stated above and setting
$E_F = 3\ \textup{meV}$ and $p_y = 2\ \textup{\AA}$, we compute the OM of all occupied states in the vicinity of the band crossing from both modern theory, Eq. (\ref{eq: two band 1}), and ACA, Eq.~\eqref{eq:atom-centerd approximation}. By replacing further $E_F$ with $|\mathbf{d}(\mathbf{k})|$ in Eq.~\eqref{eq: two band 1}, we obtain the Fermi-surface contribution due to the self-rotation of the wave packet \cite{Xiao2005}, which determines the orbital magnetoelectric response \cite{Zhong2016, Yoda2015}. Based on the results shown in Fig. 5$\textbf{b}$, we conclude that
while the values of $\mathbf{m}_\pm^\textup{ACA} (\mathbf{k})$ are strictly bound by $\mu_B$ everywhere as can be confirmed analytically, the pronounced singular behavior of $\mathbf{m}_\pm^\textup{mod} (\mathbf{k})$ with large values within the gap can be attributed solely to the 
rapid variation of the electric polarization in the vicinity of the crossing.


\

\noindent
\textbf{\large{Discussion}}

\noindent
We have shown that $sp$ orbital hybridization is the main mechanism for the ORE at surfaces of $sp$-alloys, which is manifest already without SOC. Just like the spin Rashba effect follows from the ORE via SOC, we can expect other 
orbital-dependent phenomena to be formulated and discovered, from which SOC recovers 
their spin analogues. One remarkable example is orbital analogue of the quantum anomalous Hall effect, where the quantized edge state is orbital-polarized \cite{Wu2008}. Another example is the recent formulation of the orbital version of the Dzyaloshinskii-Moriya interaction governing the formation of chiral structures such
as domain walls and skyrmions~\cite{Kim2013c}. 

Our simulations reveal the complexity of the orbital textures driven by ORE, 
and their sensitivity to the electronic structure of realistic materials. This means that 
the desired properties of the ORE, and phenomena it gives rise to, can be designed by
proper electronic-structure engineering. Moreover, by making use of both spin and orbital degrees of freedom, one can generate and manipulate  arising spin and orbital textures in $\mathbf{k}$-space. While for the case of $\textup{Bi}\textup{Ag}_2$ considered here the spin aligns collinearly to the OM in the presence of SOC, a Rashba effect with respect to the {\it total} angular momentum emerges in each of the $j=3/2$ and $j=1/2$ branches in the regime where SOC is dominant over the ORE~\cite{Park2011, Hong2015}. Based on a similar idea, an
orbital version of the Chern insulator state in a situation of very large SOC has been recently proposed~\cite{Zhang2013}.

We predict that within the ORE, in contrast to the spin Rashba effect, the magnitude of the OM in the vicinity of band crossings can reach gigantic values. This observation has very far-reaching consequences for the magnitude of effects which are directly associated with the OM at the Fermi surface. This particularly concerns the orbital magnetoelectric effect, as recently discussed 
in the context of orbital Edelstein effect~\cite{Yoda2015} and gyrotropic magnetic effect~\cite{Zhong2016}. Within the orbital Edelstein effect, a finite OM at the Fermi 
surface is generated by an asymmetric change in the distribution function created by an 
electric field. In the gyrotropic magnetic effect, discussed intensively these days with respect 
to topological metals~\cite{Xu2015, Lv2015}, an external magnetic field
gives rise to an electrical current~\cite{Zhong2016}. Both phenomena rely crucially on the magnitude of the local orbital moments at the Fermi surface of materials, and 
we predict here that they can be drastically enhanced by tuning the electronic structure such 
that the singularities in the OM, which we disclose in our work, are positioned at the Fermi
level. The generally enhanced predicted magnitude of the Berry-phase OM of the occupied states, as compared to the OM computed from the commonly used approximation, can also manifest in the re-evaluation of the magnitude of other effects such as the orbital Hall effect~\cite{Bernevig2005, Kontani2009}.

An additional flavor to the ORE is the intrinsic {\it valley} degree of freedom inherent
to systems of the type studied here: because of time-reversal symmetry the points of singularity in the OM always come in pairs. Since the OM is opposite for opposite valleys, 
the overall OM integrated over the Brillouin zone is zero in the ground state, and orbital magnetoelectric response becomes strongly valley-dependent. Exploiting the ORE for the purpose of generating sizable
ground state net OM at surfaces has to be done in combination with generating a non-vanishing exchange field and magnetization in the system. This can give rise to
a plethora of effects relying on intertwined spin and orbital degrees of freedom in complex
magnetic materials.

\

\noindent
\textbf{\large{Methods}}
\

\noindent
\textbf{First principles calculation.}
We performed self-consistent density-functional theory calculations of the electronic structure of $\textup{Bi}\textup{Ag}_2$ using the film mode of the \texttt{FLEUR} code \cite{Fleur}, which implements the FLAPW method \cite{Wimmer1981, Krakauer1979}. Exchange and correlation effects were treated within the generalized gradient approximation \cite{Perdue1996}. We assumed a $(\sqrt{3}\times\sqrt{3})R30^\circ$ unit cell (see Fig. 2$\textbf{a}$) with the in-plane lattice constant $a=9.47\,a_0$, where $a_0$ is Bohr's radius. The surface relaxation of Bi was set to $d=1.61\,a_0$, and the muffin-tin radii of Bi and Ag were chosen as $2.80\,a_0$ and $2.59\,a_0$, respectively. We used $K_\textup{max} = 4.0\,a_0^{-1}$ as plane-wave cutoff and sampled the irreducible Brillouin zone using $110$ points. Spin-orbit coupling was included self-consistently within the second-variation scheme \cite{Li1990}.

Based on the converged charge density, maximally-localized Wannier functions (MLWFs) were obtained in a post-processing step employing an equidistant $16\times 16$ $\mathbf k$-mesh. Starting from $sp_2$ and $p_z$ trial orbitals on Bi as well as $s$, $p$, and $d$ trial functions on Ag, we constructed 44 MLWFs out of 120 energy bands using the \textsc{wannier}{\footnotesize{90}} program~\cite{Wannier90}. The frozen window was set $2.78\ \textup{eV}$ above the Fermi energy. Subsequently, we calculated the OM in (i)~the ACA, and (ii)~the Berry phase theory according to the scheme proposed by Lopez \emph{et al.} \cite{Lopez2012}.

\

\noindent
\textbf{\large{Acknowledgment}}
\

\noindent
D.G. acknowledges the financial support from the Global Ph.D. Fellowship Program funded by NRF (2014H1A2A1019219). H.W.L is supported by the National Research Foundation of Korea (NRF) grant  (No. 2011-0030046). We thank Juba Bouaziz, Manuel dos Santos Dias, Philipp R\"u\ss mann, and Samir Lounis for stimulating discussions. 


\

\noindent
\textbf{\large{Author contributions}}
\

\noindent
D.G. uncovered the mechanism of the orbital Rashba effect by tight-binding model consideration. D.G., J.-P.H., and P.M.B. performed first principles calculations. D.G., J.-P.H., and Y.M. wrote the manuscript. All authors discussed the data and contributed to the paper.

\

\noindent
\textbf{\large{Competing financial interests}} 

\noindent
The authors declare no competing financial interests.

\end{document}